Morphology of graphene thin film growth on SiC(0001)

Taisuke Ohta[1,2], Farid El Gabaly[3], Aaron Bostwick[1], Jessica McChesney[1,2], Konstantin V. Emtsev[4], Andreas K. Schmid[3], Thomas Seyller[4], Karsten Horn[2], Eli Rotenberg[1]

[1]Advanced Light Source, Lawrence Berkeley National Laboratory, Berkeley, California, USA

[2]Fritz-Haber-Institut der Max-Planck-Gesellschaft, Berlin, Germany

[3]National Center for Electron Microscopy, Lawrence Berkeley National Laboratory, Berkeley, California, USA

[4]Lehrstuhl für Technische Physik, Universität Erlangen-Nürnberg, Erlangen, Germany

Abstract

Epitaxial films of graphene on SiC(0001) are interesting from a basic physics as well as applications-oriented point of view. Here we study the emerging morphology of *in-vacuo* prepared graphene films using low energy electron microscopy (LEEM) and angle-resolved photoemission (ARPES). We obtain an identification of single and bilayer of graphene film by comparing the characteristic features in electron reflectivity spectra in LEEM to the π-band structure as revealed by ARPES. We demonstrate that LEEM serves as a tool to accurately determine the local extent of graphene layers as well as the layer thickness.

Graphene, the single layer building block of graphite and carbon nanotubes, possesses a number of properties that set it apart from other materials, such as the massless Dirac fermion character of the charge carriers that is associated with the linear dispersion of the electronic bands near the Fermi level. This results in intriguing transport properties[1] that render graphene an interesting material for basic physics studies of a model 2D solid, and for potential applications in electronic devices. While transport studies so far have concentrated on exfoliated "flakes" of graphene on oxidized silicon wafers with a size of ~100 μm [2,3], practical applications in electronics require large scale epitaxial films. One route towards achieving this goal consists of removing the silicon atoms in the top layer(s) of a silicon carbide, SiC(0001) substrate through annealing[4,5], leaving behind an electronically decoupled epitaxial graphene layer[5,6,7]. Such films have been the basis for electronic structure studies of single and multiple layer graphene films [6,7,8,9,10,11,12]. However, little is known about the film morphology that results from this preparation process, such as how to optimize the growth conditions for large-scale uniformity, or the details about the interface between graphene and SiC. Here we report on a combined low energy electron microscopy (LEEM) and angle-resolved photoemission (ARPES) study of graphene films grown on SiC(0001) that reveals the structure of the as-prepared films, and permits an unambiguous correlation of features in the electron reflectivity spectra from LEEM to specific film thicknesses at the various stages of film growth.

LEEM experiments were carried out at the National Center for Electron Microscopy at Lawrence Berkeley National Laboratory, in a custom-built instrument[13,14] with a base pressure of 3 x 10[-11] mbar, using the bright field mode of operation (i.e., using the specularly reflected electron beam for imaging). Clean SiC samples were prepared by the method described in Ref.15, and were transferred into the chamber through a load lock. The sample temperature was then raised up to the required temperature, as determined from a thermocouple connected to the sample holder close to the specimen[16], for no longer than 30 seconds. This resulted in the substrate surface being converted to graphene layers of different thickness varied by choice of the annealing temparature.[6] LEEM images were recorded at specific electron energies. They revealed strong contrast variations due to quantum size effects in the band structure above the vacuum level, $E_{vac}$, which influence sample reflectivity[17] as discussed in more detail below. All electron reflectivity spectra were acquired at room temperature.

A LEEM image recorded after annealing the SiC to 1600C is shown in Fig. 1 (a). The electron energy of $E_{vac}$ + 4.88 eV was used, and the field of view (FOV) is 8 μm. There are several distinct regions with up to micrometer linear dimensions and with contrast revealed in four different shades of grey. The image is part of a series of 128 LEEM images of the same region, recorded as a function of electron energy, varied from -1 to 12 eV with respect to $E_{vac}$. This set of images was used to extract spatially resolved LEEM image intensity spectra from the different regions of the graphene film. Figure 1 (b) shows the electron reflectivity[17] as a function of electron energy above $E_{vac}$, extracted along the white dashed line indicated in Fig. 1 (a). The electron reflectivity in the image (Fig. 1(b)) has three representative patterns; these are extracted in Fig. 1(c). Line-colors of the plotted spectra are matched to the grey scale of the image in Fig. 1(a), except for the top and bottom spectrum (dashed and dash-dot-dash lines).

The electron reflectivity spectra of graphene layers have been previously recorded by Hibino *et al.*,[18] and were analyzed in terms of band structure calculations. These authors concluded that the low reflectivity just above $E_{vac}$ corresponds to the energy position of the conduction band in graphite, which splits along the wavevector perpendicular to the graphite's 2D plane (Γ to A point) and becomes discrete as the bands are quantized due to the finite thickness. The spectra recorded in this study closely match the data of Hibino *et al*. Furthermore, the corresponding thickness for each reflectivity spectrum, where determined by comparing the LEEM and ARPES experiments, is discussed in the following section.

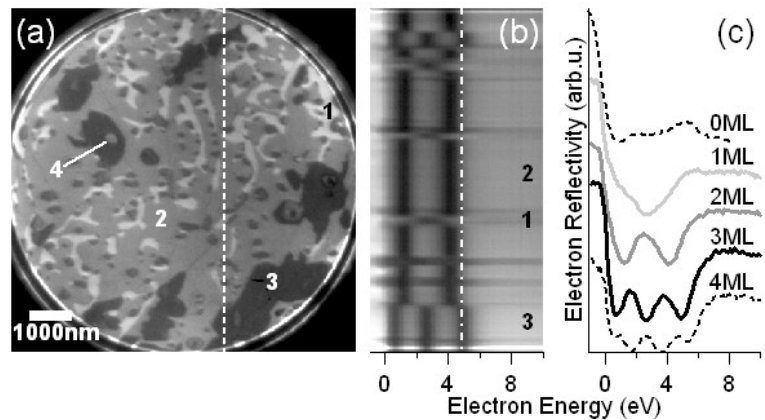

Figure 1: (a) LEEM image of a graphene film on SiC(0001), recorded with the electron energy of $E_{vac}$ + 4.88 eV and 8 μm field of view. (b) Intensity as a function of electron energy for a section along the white dashed line in (a) is represented on a grey-scale, revealing features associated with areas of different reflectivity (brightness) in the image (a). The electron energy is measured with respect to $E_{vac}$, and the electron energy of (a) is indicated as the white dash-dot-dash line. The numbers superimposed in (a) and (b) correspond to the number of electronically active layers. (The procedure we use to identify layer thickness by comparison with ARPES results is described in the text) (c) Electron reflectivity spectra of five representative regions corresponding to the number of layers, where 0ML corresponds to the electronically inactive buffer layer, and one, two, three and four graphene layers.

LEEM images after two successive annealing cycles (1280C and 1310 C) demonstrate the early stage evolution of graphene formation at elevated temperatures (Figs. 2(a) and (b)). Because the sample had to be moved away from the LEEM electron optics during annealing, these images represent similar but not identical regions on the surface, and they are representative of the results of annealing steps at different temperatures. Both images exhibit three distinct intensities at the same electron energy.

We now use images such as these to perform a "population analysis" of the different patches, and to confirm the identification of the patches' layer thickness through comparison with the ARPES measurement. The shapes of the light and medium grey regions appear fairly irregular; however, the location of the light grey region seems to follow the pattern of the atomic step structure of the original SiC substrate (not shown). While the medium grey regions are dominant in the images (a) and (b), the light grey regions are reduced and the black regions become larger in (b). The electron reflectivity spectra from the light and medium grey and black regions are very similar to those labeled 0ML, 1ML and 2ML in Fig. 1 (c).

This observation is quantified in Fig. 2 (e), which is the histogram of the surface area for the different regions before and after the annealing step. By removing the overall background due to the inhomogeneous beam shape and multichannel plate, and applying an object extraction process,[19] the Fig. 2 (a) and (b) are divided into regions of three intensity levels as shown in Fig. 2 (c) and (d). The area fraction of the light grey regions (0ML in Figs. 2 (a) and (b)) was reduced through the annealing cycle, while the area fraction of the black regions (2ML) increased, resulting in the overall increase of the graphene thickness.

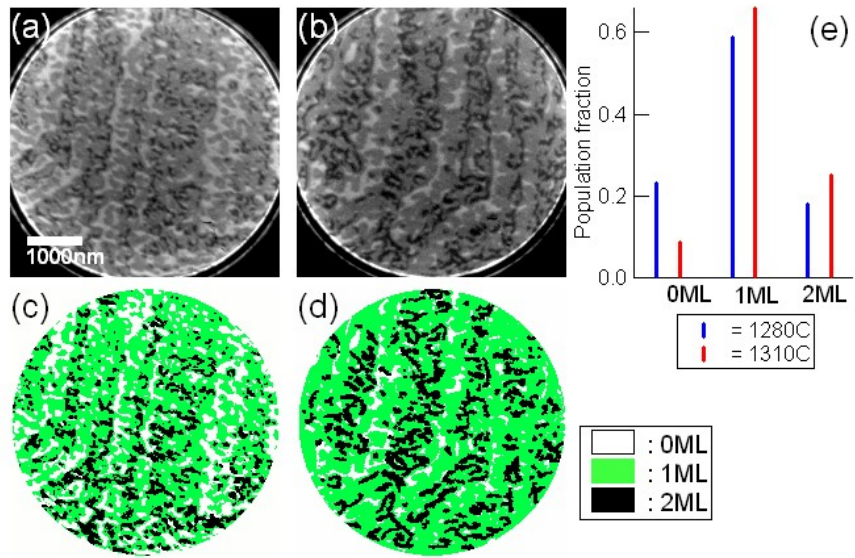

Figure 2: (a) and (b) (color online) LEEM images of the early stage of graphene formation after annealing at 1280 and 1310 C, and field of view 5 μm, recorded with the electron energy of $E_{vac}$ + 4.08 eV. (c) and (d) Composite images of three different regions extracted through an object extraction process on (a) and (b), respectively. (e) Histogram of the surface area for the different regions before and after the annealing step. Bars on the left from (c), and ones on the right from (d).

Using LEEM, the evolution of film morphology can be studied under a greatly varying lateral scale, ranging from nanometers to micrometers [20], and usually regions with different thickness or structure can be imaged with good contrast. To confirm unambiguous interpretation of these images in terms of local film thickness and structure, we use additional information obtained by ARPES. As shown in a series of studies[6,7,], ARPES provides a means of identifying the thickness of the majority species in the film, not only through a method that is laterally resolving, but also is sensitive to thickness in a "fingerprint" manner. A single linearly dispersing π band at the K point of the Brillouin zone, near the so-called Dirac point $E_D$ [2,3], is characteristic of the single layer, while the multilayer spectrum has an extra set of bands due to the interlayer splitting brought about by the close proximity of the layers. Thicker films have a consecutively higher number of bands, which can clearly be distinguished for up to four layers [11].

One of the intriguing scientific questions in graphene films grown on SiC is the structure of the interface carbon-rich layer between the substrate and the decoupled graphene film, which exhibits a complex (6√3 x 6√3) LEED pattern. This carbon-based interfacial layer, which so far has not been well characterized structurally, is thought to serve the important function of saturating the substrate dangling bonds, thus electronically decoupling the graphene layer from the substrate.[5,21,22] However, it has been argued that the 6√3 x 6√3 structure arises as a Moiré pattern from the mismatch between the substrate and graphene lattice, such that this structure must be considered as the first graphene layer[23].

In order to perform such a comparison, samples that had been characterized by LEEM were transported to the Advance Light Source, Lawrence Berkeley National Laboratory, and inserted into the Electronic Structure Factory endstation on beamline 7.[6,7] Graphene films are stable in air even after prolonged periods of storage, apart from adsorption of water, which can be readily desorbed after a mild anneal of the sample in vacuum in the $10^{-11}$ mbar range. After this procedure, no contaminants such as oxygen or nitrogen were observed in the detection limit of core level photoemission at the photon energy of 730 eV.

ARPES data as a function of energy and electron wavevector, shown in Fig. 3(a), were recorded at a photon energy of 95 eV, near the K point of the Brillouin zone, along the M-K-Γ line (from left to right in Fig. 3(a)), from the sample shown in Fig. 2(b). These data reveal features due to single and bilayer regions of the film as follows. While the single layer is characterized by a linear dispersion as indicated by the medium grey lines in Fig. 3(b), the double layer exhibits two sets of bands due to interlayer interaction.[6,11,12] Three separate bands are observed in the measured photoemission pattern (Fig. 3(a)), one from the monolayer band, and two from the branches of the bilayer bands.  The assignment of monolayer and bilayer bands is based on our previous ARPES results from ref. 11.  Notice that only one side of the bands, along Γ-K, is visible due to a photoemission cross section effect.[24]

A section through the photoemission pattern along the dashed line in Fig. 3(a) shows the photoemission intensity for each band (Fig. 3 (c)). These lines were subjected to a Lorentzian line-shape fit in order to extract band intensities that are then related to the respective fraction of each layer on the surface. Since the bands from bilayer graphene are split, the intensities of the two bands are added to compare them to the single layer.[25]  The area fractions are averaged over the energy range between $E_F$-1.5 eV to $E_F$-2.0 eV, and plotted in Fig. 3(d) together with the area fractions determined from the LEEM measurement.  Twelve randomly chosen regions of ~5 μm FOV LEEM images (not shown) were examined to quantify the respective area fractions from samples identical to that in Fig. 2 (b).  Since the interface layer displays only a diffuse background intensity near $E_D$ [5,7,26] in ARPES and the population fraction for the interface layer of this sample is very small, we limit our analysis to single and double layer assignments.  While with LEEM, we can identify three different kinds of regions in Fig. 2 (b).  The area fractions for single and double graphene layers obtained from ARPES show close agreement with the area fraction for medium grey and black regions, after subtracting the area fraction for light grey (0 ML) regions (see the scales are different for LEEM and ARPES measurements in Fig. 3 (d)). This permits a direct assignment of the electron reflectivity spectra from the medium grey and black regions in the LEEM images (Fig. 2 (a) and (b)) to single and double graphene layers, and we can propose that the light grey ('0 ML') regions in LEEM are the carbon-rich interface layer. Moreover, from the histograms in Fig. 2 (e), we confirm the reduction of interface layer regions and the increase of double graphene layers by increasing the annealing temperature a small amount.

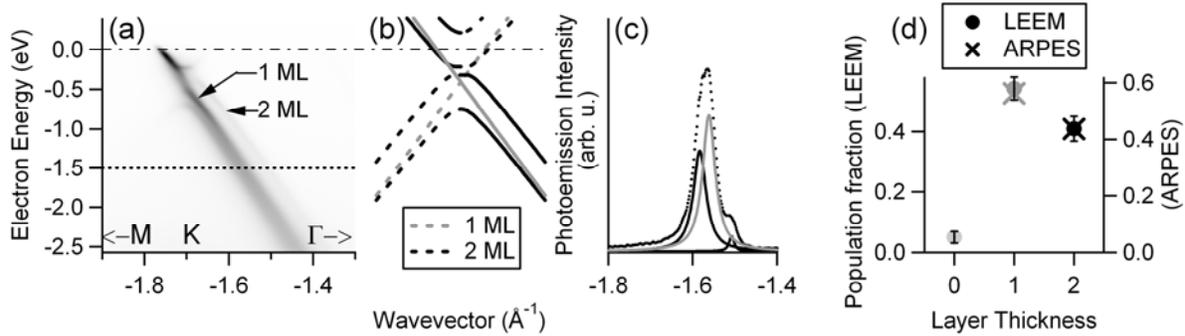

Figure 3: (a) Photoemission intensity pattern of the graphene film shown in Fig. 2 (b), recorded at a photon energy of 95 eV in the region near the K point in the Brillouin zone. The lines in grey scale in (b) indicate the tight binding band structures from single and double graphene layers [11,12,27,28]. (c) A section through the photoemission pattern along the dashed line in (a) with the Lorentzian line-shape fitted functions assigned to single (medium grey line) and double (black lines) graphene layers. (d) Respective area fractions determined through LEEM and ARPES experiments on the same sample. Note that the vertical scale for ARPES is scaled to account for the lack of the interface carbon-rich layer component, so that LEEM and ARPES results can be directly compared. The error bars in the LEEM results represent the standard deviations of the respective area for each regions in twelve images examined.

By comparing LEEM images and the valence structures from ARPES, we have identified the electron reflectivity spectrum of the interface carbon-rich layer (top spectrum in Fig. 1 (c)), which is qualitatively different from those of graphene layers. We postulate that the low electron reflectivity at higher electron energy (>$E_{vac}$ + 6eV) is due to the smaller band dispersion for the unoccupied band structure of the interface carbon-rich layer, which are also seen in its occupied states[5].  This strongly suggests that the interface layer cannot be considered as the first electronically active layer in accord with earlier valence band studies[5] and recent theoretical studies[21,22].

In summary, we have monitored the evolution of graphene layers on SiC(0001) using LEEM. On the basis of electron reflectivity spectra, calibrated using ARPES, we arrive at an unambiguous assignment of regions in the LEEM images due to the interfacial layer and single-layer as well as bilayer graphene. This information can serve as the basis for an improvement of layer morphology of large scale epitaxial graphene films through optimization of preparation condition.

Acknowledgements

We are grateful to H. Hibino and M. Nagase at NTT Basic Research Laboratories for fruitful discussions, and sharing their results prior to publication.  This work and the Advanced Light Source were supported by the U.S. Department of Energy, Office of Basic Sciences under Contract No.DE-AC02-05CH11231.  T.O. and J.L.M. gratefully acknowledge support from Lawrence Berkeley National Laboratory and the Max Planck Society, and F. E.-G. from the Comunidad Autonoma de Madrid and Universidad Autonoma through project No. CCG06-UAM/MAT-0364.

[1] A.K.Geim and K.Novoselov, Nature Materials **6**, 183 (2007).

[2] K. S. Novoselov *et al.*, Nature **438**, 197 (2005).

[3] Y. Zhang *et al.*, Nature **438**, 201 (2005).

[4] I. Forbeaux, J.-M. Themlin, and J.-M. Debever, Phys. Rev. B **58**, 16396 (1998).

[5] K. Emtsev *et al.*, Mat. Sci. Forum 556-557, 525 (2007).

[6] T.Ohta, A.Bostwick, Th. Seyller, K.Horn, E.Rotenberg, Science **313**, 951 (2006).

[7] Aaron Bostwick, Taisuke Ohta, Thomas Seyller, Karsten Horn, Eli Rotenberg Nature Physics, **3**, 36 (2007).

[8] Th. Seyller *et al.*, Surf.Sci **600**, 3906 (2007).

[9] E. Rollings et al., J. Phys. Chem. Solids **67**, 2172 (2006).

[10] G. M. Rutter, J. N. Crain, N. P. Guisinger, T. Li, P. N. First, J. A. Stroscio, Science **317**, 219 (2007).

[11] T.Ohta, A. Bostwick, J.L.McChesney, Th. Seyller, K.Horn, E.Rotenberg, Phys. Rev. Lett. **98**, 206802 (2007).

[12] S. Y. Zhou *et al.*, Nature Materials, published online: 09 September (2007).

[13] K. Grzelakowski, T. Duden, E. Bauer, H. Poppa, S. Chiang, IEEE Transactions on Magnetics **30**, 4500 (1994).

[14] T. Duden and E. Bauer, Surf. Rev. Lett., **5**, 1213 (1998).

[15] Th. Seyller, Applied Physics A-Materials Science & Processing **85**, 371 (2006).

[16] The thermocouple was calibrated using an infrared pyrometer with an emissivity setting of 0.65.

[17] E. Bauer, Rep. Prog. Phys. **57**, 895 (1994).

[18] H. Hibino, H. Kagashima, F. Maeda, M. Nagase, Y. Kobayashi, H. Yamaguchi (2007) (arXiv:0710.0469).

[19] The object extraction was performed by thresholding of grey scale images to two different intensity levels to distinguish the light and medium grey and black regions, followed by erode-dilate cycles of binary images to reduce image noise. This procedure gave equivalent population distribution as a fit to the grey scale histogram to three Gaussian functions of equal width.

[20] W. Telieps and E. Bauer, Surf. Sci. **162**, 163 (1985).

[21] F.Varchon *et al.*, Phys. Rev. Lett. **99**, 126805 (2007).

[22] A. Mattausch and O. Pankratov, Phys. Rev. Lett. **99**, 076802 (2007).

[23] M. H. Tsai *et al.*, Phys. Rev. B **45**, 1327 (1992); L. Li and I. S. T. Tsong, Surf. Sci. **351**, 141 (1996); L. Simon, J. L. Bischoff, L. Kubler, Phys. Rev. B **60**, 11653 (1999).

[24] Shirley, E., Terminello, L., Santoni, A. & Himpsel, F. J., Phys. Rev. B **51**, 13614–13622 (1995).

[25] The relative intensity of the two bands of bilayer graphene is a function of the photon energy used for ARPES measurement.  See ref. 11.

[26] A. Bostwick *et al.*, accepted to New J. Phys. (2007) (arXiv:0705.3705).

[27] F. Guinea, A. H. Castro Neto, N. M. R. Peres, Phys. Rev. B 73, 245426 (2006).

[28] E. McCann and V. I. Fal'ko, Phys. Rev. Lett. **96**, 086805 (2006).